\documentclass[aps,prl,reprint,superscriptaddress,floatfix]{revtex4-2}

\usepackage{graphicx}
\usepackage{amsmath,units}
\usepackage{mathrsfs}
\usepackage{color} 
\usepackage{xcolor} 
\usepackage[export]{adjustbox}
\usepackage{ulem}

\begin{document}

\title{Piston-Like Information Engine II: Boundary-Controlled Optimum in Active Matter}

\author{Laura Hoek}
\affiliation{School of Chemistry, Raymond and Beverly Sackler Faculty of Exact Sciences, Tel Aviv University, Tel Aviv 6997801, Israel}
 \author{Neta Ben Ari}
\affiliation{Schulich Faculty of Chemistry, Technion-Israel, Institute of Technology, Haifa 3200003, Israel}
\author{R\'emi Goerlich}
\affiliation{School of Chemistry, Raymond and Beverly Sackler Faculty of Exact Sciences, Tel Aviv University, Tel Aviv 6997801, Israel}

 \author{Saar Rahav}
\email{rahavs@ch.technion.ac.il}
\affiliation{Schulich Faculty of Chemistry, Technion-Israel, Institute of Technology, Haifa 3200003, Israel}
\author{Yael Roichman}
\email{roichman@tauex.tau.ac.il}
\affiliation{School of Physics \& Astronomy, Raymond and Beverly Sackler Faculty of Exact Sciences, Tel Aviv University, Tel Aviv 6997801, Israel}
\affiliation{School of Chemistry, Raymond and Beverly Sackler Faculty of Exact Sciences, Tel Aviv University, Tel Aviv 6997801, Israel}
\affiliation{Center for the Physics and Chemistry of Living Systems. Tel Aviv University, 6997801, Tel Aviv, Israel}

\date{\today}

\begin{abstract}
Information engines convert information into extractable work through measurements and feedback. We realize an engine that employs information to compress a gas of self-propelled bristle bots. Its work per measurement, $\bar{W}$, is controlled by the size of the detection region $\Delta x$ and the probability $p_1$ that this region is vacant. In thermal systems, $\bar{W}$ follows the universal form $-p_1 \ln p_1$, whereas the active engine shows a qualitatively modified relation. As particle density increases, the maximum of $\bar{W}(\Delta x)$ switches between two distinct operating regimes. Notably, this transition is also found in the solutions that maximize power output for finite-time cycles with dynamics affected by dry friction. We attribute this nonequilibrium feature to the accumulation of active particles near the boundaries.  
\end{abstract}

\maketitle

Information engines are devices that use measurements and feedback to convert fluctuations into work. Szil\'ard proposed the first model of such an engine~\cite{szilard1929entropieverminderung, szilard1964decrease} to quantify the effects of Maxwell's demon~\cite{Maxwell_book}. Renewed interest in the connections between information and thermodynamics~\cite{parrondo_thermodynamics_2015} led to the generalizations of the second law~\cite{sagawa_generalized_2010,horowitz_nonequilibrium_2010}, as well as investigations of the cost of logical operations~\cite{landauer1961irreversibility, berut_experimental_2012}. Advances in miniaturization and control have since allowed for a variety of experimental realizations of information engines~\cite{toyabe_experimental_2010, koski_experimental_2014-1,paneru_lossless_2018,admon_experimental_2018,lee_experimentally_2018,paneru_optimal_2018,ribezzi-crivellari_large_2019,paneru_efficiency_2020,saha_maximizing_2021, Saha2023Information}.

In a companion study, an information engine used feedback to compress colloidal particles with a piston formed by optical traps~\cite{remi_now}. Periodic measurements assessed whether the detection region near the piston was empty, thereby enabling near-work-free compression. It was found that for thermal particles with hard walls, the thermodynamics of the information engine is characterized by the probability that the observed area is empty, $p_1$. Specifically, the mean work stored per measurement was found to be 
\begin{equation}
\bar{W} =-k_{\mathrm{B}}T p_1\ln{p_1} .
\label{eq:eq_Wbar}
\end{equation}
$\bar{W}$ has a maximum at $p_1=1/e$, for any particle density or interparticle interactions. The condition $p_1=1/e$ can be viewed as a criterion for choosing a measurement and compression step that maximizes the energetic impact of each measurement.

The study of active matter, collections of self-propagating particles, has attracted considerable interest~\cite{vicsek2012collective,Solon2024,di_leonardo_bacterial_2010,Aranson_2022}. Active particles exhibit diverse phenomenology that includes flocking~\cite{vicsek2012collective, Liebchen2017}, phase separation~\cite{cates2015motility}, and large fluctuations~\cite{Deseigne2010}. Due to their self-propulsion, active particles are inherently out-of-equilibrium, and a thermodynamic interpretation of their motion is nontrivial~\cite{Solon2015}. Several recent works have explored information engines with active matter~\cite{Paneru2022Colossal, Malgaretti2022Szilard, Omer_2023, Saha2023Nonequilibrium, Cocconi2025Mechanical}, raising the question of how the difference between active and thermal working media manifests in the operation of such engines.
 \begin{figure}[b]
    \centering
    \includegraphics[width=1\linewidth]{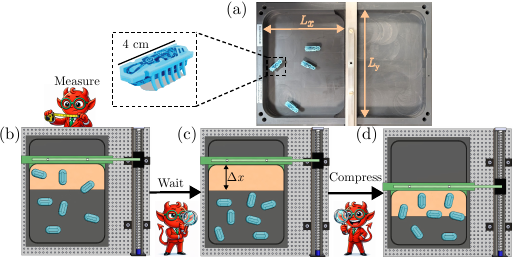}
    \caption{\textbf{Experimental setup and piston-like working cycle.}
(a) Rectangular arena containing active particles (bbots); inset: a single bbot. A linear actuator controls the movable boundary. Bottom: piston-like working cycle.
(b) The number of bbots in the detection region, $\Delta x L_y$, is measured periodically ($L_y=29.2$\thinspace{cm}).
(c) If the region is occupied, the boundary remains fixed.
(d) If empty, the boundary moves inward, compressing the bbots. Measurements are spaced sufficiently far apart to be uncorrelated.}
    \label{fig: PLIE schematics}
\end{figure}

Here, we experimentally realize a macroscopic information engine with an active working medium. Its operating principle closely follows that of Ref.~\cite{remi_now}, allowing a direct comparison that highlights the role of particle activity. We argue that, in this class of engines, persistent activity manifests as deviations of the mean work per measurement from $-p_1 \ln p_1$.  We attribute these deviations to the accumulation of active particles near the piston boundaries, a phenomenon previously studied in other contexts~\cite{elgeti2013wall, Duzgun2018ABPWalls, Wagner2022SteadyStates}. Our active information engine exhibits intricate behavior in which the maximum of $\bar{W}(p_1)$ switches between operating schemes that use large detection regions and those that probe regions within the accumulation layer.

Our setup is shown schematically in Fig.~\ref{fig: PLIE schematics}. A rectangular arena of dimensions $L_x \times L_y$, with rounded corners and area $A$, is fitted with a moving wall, forming a two-dimensional piston. $N_b$ bbots are placed inside, and their motion is tracked using a high-resolution webcam (Logitech Brio at $30$\thinspace{fps}) and standard particle tracking algorithms. At time intervals $t_m$, a detection region of width $\Delta x$ and area $L_y \Delta x$ near the movable wall is examined. The time interval $t_m$ is chosen to be larger than all relevant correlation times, ensuring that measurements are always uncorrelated. If the region is found to be empty, with probability $p_1 (\Delta x)$, the wall is moved inwards by $\Delta x$. Nothing is done if one or more bbots are found in the observation area. Repeatedly applying this process compresses the bbot gas without performing direct work. Engine performance can be optimized by adjusting the size of the detection region.

The engine's thermodynamic characterization has mechanical and informational components. The relevant mechanical quantity is the work stored in each compression step, which depends on the force $F=P L_y$ exerted by the bbots on the piston, where $P$ is the two-dimensional pressure. A compression from $L_x$ to $L_x-\Delta x$ stores work of $W_{\mathrm{st}} = \int_{L_x-\Delta x}^{L_x} P(x) L_y dx$. The mean work per measurement is therefore $\bar{W}=p_1 W_{\mathrm{st}}$.

\begin{figure}[t]
    \centering
    \includegraphics[width=1\linewidth]{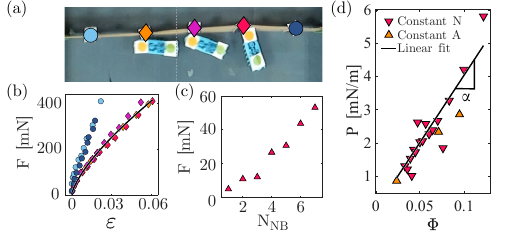}
    \caption{\textbf{Experimental pressure measurements.}
(a) Photograph of the experimental setup with the flexible boundary; markers indicate the positions used for force calibration.
(b) Force--strain calibration curves measured at the positions indicated in (a).
(c) Force exerted by boundary-localized clusters of $N_{\mathrm{NB}}$ bbots at fixed arena area, inferred from the measured boundary deflection.
(d) Pressure $P$ versus area fraction $\Phi$. The black line is a linear fit with slope $\alpha$. Red: fixed particle number with varying area; orange: fixed area with varying particle number.}
    \label{fig: pressure graphs}
\end{figure}

The interaction between bbots and the walls involves repeated weak collisions, making it hard to measure the forces they apply directly. We estimated the force using the calibration method depicted in Fig.~\ref{fig: pressure graphs}.
Following~\cite{Junot2017}, we replace the rigid wall with a flexible nylon barrier reinforced by vertical supports [Fig.~\ref{fig: pressure graphs}(a)] to capture the cumulative force of multi-particle collisions. We calibrate the partition's stiffness by applying known forces $F$ via a force gauge and measuring the resulting strain, $\varepsilon = \frac{L}{L_0}-1$. The results can be fitted to a power law, $F=\mu \varepsilon^\sigma$, with $\mu=3000\pm 300 \; \mathrm{mN}$ and $\sigma=0.71\pm0.03$. This force-strain calibration curve [Fig.~\ref{fig: pressure graphs}(b)] is consistent across the soft barrier, except near the clamped edges where edge effects occur. Next, we measured the deflection of the flexible wall as a function of the number $N_{\text{NB}}$ of bbot in contact with the wall. The calibration curve in Fig.~\ref{fig: pressure graphs}(b) is then applied to express the force as a function of $N_{\text{NB}}$, shown in Fig.~\ref{fig: pressure graphs}(c).  A linear dependence is found.

Finally, during the experiment, the mean number of bbots in contact with the piston is determined by tracking their trajectories in the arena. The curve depicted in Fig.~\ref{fig: pressure graphs}(c) is used to determine the relevant pressure.
Fig.~\ref{fig: pressure graphs}(d) shows the resulting two-dimensional pressure as a function of the bbots area fraction $\Phi=N_b A_b / L_x L_y$. Here $A_b$ is the area occupied by a single bbot. Measurements for fixed $N_b$ and varying $L_x$, as well as measurements with fixed area and varying $N_b$, were found to collapse onto the linear relation
\begin{equation}
P = \alpha \Phi,
\label{eq:_Pressure}
\end{equation}
with $\alpha = 40.3 \pm 3.7$\thinspace{mN$/$m} [Fig.~\ref{fig: pressure graphs}(d)].
The resulting stored work is given by
\begin{equation}
W_{\mathrm{st}} = -\tilde{\alpha} N_b \ln(1 - \delta x),
\label{eq:Wout_log}
\end{equation}
where $\delta x = \Delta x / L_x$, and $\tilde{\alpha} = \alpha A_b = 2.0 \pm 0.2$\thinspace{mN$\cdot$cm}.


The linear relation between force and density implies that the mechanical properties of the engine are analogous to those of an ideal-gas counterpart, with an energy scale of $\tilde{\alpha}$ instead of $k_B T$. 
The former is larger by~15 orders of magnitude, reflecting the macroscopic nature of bbots. Its magnitude, $\tilde{\alpha}\sim10^{-5}$\thinspace{J}, is also consistent with effective energy scales reported for other centimetric bbot systems~\cite{boriskovsky2026probing} and fluidized granular gases~\cite{abate2008effective}.
This difference in energy scale will not affect the comparison below, as work is properly rescaled for both active and thermal engines.

Deviations from linearity in Eq.~(\ref{eq:_Pressure}) are expected at sufficiently high densities, due to exclusion interaction and the tendency of collisions to align elongated bbots. Within the density range explored here, however, no such deviations were identified. 
Previous studies of strongly confined bbots \cite{Leoni2020Hexbug, Horvath2023Hexbug, Neta_and_Saar} suggest that the force applied on the wall may depend on its velocity or the inelasticity of collisions. These possible sources for differences between the flexible nylon and hard walls are not relevant for the comparison between active and thermal engines done here, since the dependence on $\alpha$ is eliminated by rescaling. 

\begin{figure}[t]
    \centering
    \includegraphics[width=1\linewidth]{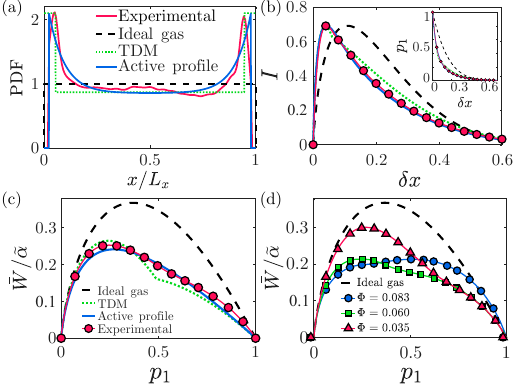}
    \caption{\textbf{Thermodynamic properties of the information engine.} (a) One-dimensional projection of the bbots density, compared to the density of an ideal gas, and two simple models of the boundary layer. (b) Inset: calculated probability $p_1$ to find a detection region of width $\Delta x$ empty. Panel: the Shannon information per measurement. (c) The mean work per measurement, $\bar{W}/\tilde{\alpha}$, as a function of $p_1$. Panels (a-c) correspond to experiments with area fraction $\Phi=0.073$. (d) Comparison of several $\bar{W} (p_1)$ curves, obtained for different area fractions, to $\bar{W}/k_BT=-p_1 \ln p_1$, expected for thermal particles. }
    \label{fig:single_step_measurement}
\end{figure}

The information-related characteristics of the engine result from the bbots' spatial distribution. This steady-state distribution was measured for various arena sizes and bbot numbers. A normalized one-dimensional projection of the bbot density is shown in Fig.~\ref{fig:single_step_measurement}(a) (Red curve, $\Phi=0.073$, and SI~\cite{deconvcomment}). The two peaks near the piston and the opposing wall result from the tendency of bbots to spend more time near the walls. This is a known property of active particles with a persistence time that exceeds or is comparable to the inverse rotational diffusion coefficient~\cite{elgeti2013wall, Duzgun2018ABPWalls, Wagner2022SteadyStates}. 
The probability of finding an empty detection region, $p_1 (\Delta x)$, is calculated from this measured density profile, assuming that the positions of different bbots are uncorrelated. A direct trajectory-based estimate of $p_1$, which does not rely on this assumption, is used as a consistency check and discussed in SI~\cite{SM_leftmost_estimate}. The mean information acquired by each measurement is then $I=-p_1 \ln p_1 - (1-p_1) \ln (1-p_1)$. The results are shown in Fig.~\ref{fig:single_step_measurement}(b).

We compare our results with those predicted by simple models of particle distribution. An ideal gas of thermal particles, with a uniform density profile [dashed lines in Fig.~\ref{fig:single_step_measurement}], gives the exact empty region probability $p_1=\left(1-\delta x\right)^{N_b}$.
The remaining models capture the non-uniform active particle density profile. The blue line in Fig.~\ref{fig:single_step_measurement} assumes maximal density near the wall (taking into account finite particle size), with exponential decay towards a lower bulk density. This description is motivated by results from the so-called active Brownian particle model~\cite{Duzgun2018ABPWalls, Wagner2022SteadyStates}. The green line in Fig.~\ref{fig:single_step_measurement} assumes a layer of constant high density near the walls and a lower constant density in the bulk. Both descriptions can be fitted to the measured density using only a few parameters; see SI for more details~\cite{SM_TDM, SM_ActiveProf}.

Both $p_1$ and $I$ for the bbots closely follow the corresponding results of the nonuniform models, while differing greatly from the curves calculated for the ideal gas. This shows that the engine characteristics are not very sensitive to the precise structure of the high-density layer, but are strongly affected by its existence. In particular, the high-density layer causes $p_1$ of active particles to first decrease faster, and then slower, than its ideal gas counterpart.

Next, we calculate the mean work per measurement $\bar{W}=p_1 W_{\mathrm{st}}$, and plot it as a function of $p_1$, see Fig.~\ref{fig:single_step_measurement}(c). For particles in thermalized with their environment, the mean work follows the ideal gas universal result of Eq.~\eqref{eq:eq_Wbar}~\cite{remi_now}.
Using rescaled work, we can compare this thermal prediction to the curve obtained for the active gas. We find deviations in the shape of $\bar{W}(p_1)$, the location of its maximum, and its amplitude, which we attribute here to the activity of the bbots. Figure~\ref{fig:single_step_measurement}(d) shows how these curves vary as a function of the bbot area fraction~$\Phi$. As $\Phi$ increases, the maximum shifts from $p_1^\star<1/e$ to $p_1^\star \geq 1/e$, revealing a density-dependent transition between two operating regimes.

The rich behavior of $\bar{W}(p_1)$ in Fig.~\ref{fig:single_step_measurement} stems from particle accumulation near the boundaries. This profile creates a direct trade-off between the terms in \mbox{$\bar{W}=p_1 W_{\mathrm{st}}$}. Although expanding the detection region $\Delta x$ stores more work $W_{\mathrm{st}}$, $p_1$ drops sharply across the dense boundary layer before decaying more gradually in the lower-density bulk. Consequently, two competing regimes emerge: small detection regions confined to the boundary layer (yielding high $p_1$), and large detection regions extending into the bulk (where expanding $\Delta x$ incurs only a minor reduction in $p_1$). Which family gives the global maximum is controlled by $p_1$ when $\Delta x$ becomes comparable to the accumulation layer width. At low densities, finding an empty region near the piston remains sufficiently likely that solutions with large $\Delta x$ (\textit{i.e.}, small $p_1$) maximize $\bar{W}$. As $\Phi$ increases, the empty-region probability inside the accumulation layer decreases rapidly. Consequently, the optimal regime switches to smaller detection regions confined within the high-density layer, operating at higher $p_1^\star$. This mode switching represents the primary finding of our study.

\begin{figure}[b]
    \centering
    \includegraphics[width=1\linewidth]{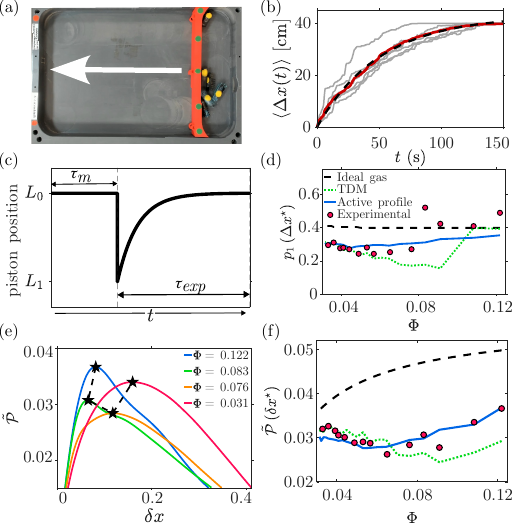}
    \caption{\textbf{Nonideal engine cycle with optimally chosen detection region}. (a) Schematic depiction of free expansion experiments. (b) Position of piston as a function of time during free expansion. Bold line depicts the mean of 7 repetitions.
    (c) Schematic of time dependence of a cycle. The piston remains at $L_0$ for several measurements; once the detection area is found to be empty, it is displaced to $L_1$, after which the system takes a time of $\tau_{\mathrm{exp}}$ to expand back to its initial volume. (d) $p_1 (\Delta x^{\star})$ as a function of area fraction $\Phi$. Symbols are calculated using experimental data, while lines are based on simple models of bbot density: Uniform (black dashed line), TDM (green dotted line), and active profile (solid blue line). 
    (e) scaled maximal power $\tilde{\mathcal{P}}$ as a function of scaled detection region size, $\delta x$, for a few representative densities. Stars mark the location of the maximum. (f) Maximum scaled power output $\tilde{\mathcal{P}}^\star$ versus area fraction. Markers and lines denote the same as panel (d).}
    \label{fig:mini_cycle}
\end{figure}

The results shown in Fig.~\ref{fig:single_step_measurement} describe an idealized engine with full control over work extraction and no friction. Can we identify a similar transition between useful engine designs with different $\Delta x$ in a nonideal operating cycle? To tackle this question, we first characterize the bbot-driven expansion of the piston. The partition was initially placed at $x_{\mathrm{min}}$, enclosing a small area, as shown in Fig.~\ref{fig:mini_cycle}(a). We then track its position as a function of time. Traces of repetitions of this expansion process can be seen in Fig.~\ref{fig:mini_cycle}(b), together with their mean (bold line). The expansion exhibits stick-slip motion, characteristic of dry friction~\cite{antonov2024inertial}.
A phenomenological fit of $\left\langle\Delta x(t)\right\rangle = \Delta X_{\mathrm{max}}(1 - e^{-t/\tau})$ captures the average expansion well. Here $\Delta X_{\mathrm{max}} = 42.10 \pm 0.04$\thinspace{cm} gives the distance that the piston can move until the end of the arena, while $\tau = 48.7 \pm 0.2$\thinspace{s} is the typical expansion time. The value of $\tau$ is expected to depend on $N_b$.

To test whether friction can account for the observed dynamics, we simulated a simplistic model of the expansion process (see SI~\cite{fricexp}). The simulations show behavior that is highly similar to the results shown in \ref{fig:mini_cycle}(b) for strong friction, but also suggest that qualitatively different behavior is possible for different parameters.

Guided by the measured expansion dynamics, we model the work-extraction cycle in four steps. First, the piston is fixed at arena length $L_x$. Second, repeated measurements test for the absence of particles within a region of width $\Delta x$. Third, upon detecting an empty region, the piston is rapidly compressed by $\Delta x$ and subsequently unlatched. Finally, the piston expands back to $L_x$ against an external force $F$, doing work during the expansion phase until it returns to its baseline configuration, completing the cycle.
The work gained in each cycle is simply 
\begin{equation}
    W_{\mathrm{out}} = F\Delta x.
    \label{eq:Wout_linear}
\end{equation}
Unlike the idealized compression studied above, this cycle has a finite duration, $\tau_c (\Delta x)$.
Although $\tau_c$ may also depend on other parameters, here we focus on its dependence on the size of the detection region. We therefore define the mean power output of the cycle as
\begin{equation}
\mathcal{P}(\Delta x,\Phi) = \frac{F \Delta x }{\tau_{\mathrm{c}}(\Delta x)}.
\label{eq:power}
\end{equation}
We look for $\Delta x$ that maximizes this output.

The time dependence of a single realization of the cycle is depicted in Fig.~\ref{fig:mini_cycle}(c). Each cycle consists of a sequence of measurements, an instantaneous compression, and a subsequent finite-time expansion. The mean cycle time is therefore
\begin{equation}
    \tau_c (\Delta x)  = \frac{t_m}{p_1 (\Delta x)} + \tau_{\mathrm{exp}} (\Delta x).
\end{equation}
Here, $t_m$ is the time between successive measurements. It is chosen to be longer than any correlation time, so that successive measurements are effectively independent. We estimated the correlation time from data and use $t_m=5 \tau_{corr}$. $1/p_1$ is the mean number of measurements before the detection region is found to be empty. For the expansion stroke, we assume that a sufficiently small $F$ does not modify the observed stick-slip dynamics. The expansion time $\tau_{\mathrm{exp}}$ can be calculated from the measured curve of~\ref{fig:mini_cycle}(b). Specifically, it is the time required for the piston to expand from $L_x-\Delta x - x_{\mathrm{min}}$ to $L_x-x_{\mathrm{min}}$. We expect this approximation to break down when $F$ is comparable to the friction force.

 The values of $\Delta x^\star$ that maximize the output, the corresponding $p_1 (\Delta x^\star)$, and the scaled maximal output $\tilde{\mathcal{P}}^\star=\mathcal{P}(\Delta x^\star) t_m/F L_x$, were computed for different area fractions. The results are shown in Fig.~\ref{fig:mini_cycle}(d)-(f).
Panel~\ref{fig:mini_cycle}(d) depicts $p_1 (\Delta x^\star)$. The experimental results are compared with predictions obtained from different models of the bbot density, including an ideal-gas-like uniform-density (dashed line). At $\Phi \simeq 0.08$, we find a transition from solutions with $p_1<1/e$ and relatively large $\Delta x^\star$ to solutions with $p_1 \geq 1/e$ and smaller $\Delta x^\star$. 
This transition is qualitatively similar to the one found in the ideal compression considered earlier.
We note that because the extracted work has a different dependence on $\Delta x$ in the present cycle, the value of $p_1$ that maximizes the output for uniform density is no longer $1/e$.

Fig.~\ref{fig:mini_cycle}(e) shows a few representative $\tilde{\mathcal{P}}$ vs $\Delta x$ curves for different area fractions. The two curves with $\Phi < 0.08$ are maximal at larger values of $\Delta x$ compared to those of $\Phi>0.08$. The value of the maximal output is shown in~\ref{fig:mini_cycle}(f). The results from panels (e) and (f) show that the output of the low probability solutions decreases with increasing $\Phi$ up to the switching point. At larger area fractions, the other family of solutions takes over, and the output starts to increase with $\Phi$. Overall, Fig.~\ref{fig:mini_cycle} shows that the trade-off governing engine performance in the ideal work extraction limit remains equally central under finite-time cycle conditions.

In conclusion, we have realized a macroscopic information engine operating on an active many-body working substance and have compared it directly with a piston-like engine operating on a thermalized working substance~\cite{remi_now}. The active gas remains mechanically ideal-like: its pressure is linear in area fraction, and the stored compression work retains the logarithmic form of the thermal case, with $k_{\mathrm B}T$ replaced by the active energy scale $\tilde{\alpha}$. Thus, activity modifies the mechanical response only through the energy scale. In addition, activity reshapes the feedback statistics that govern engine operation, manifesting as a deviation from Eq.~\eqref{eq:eq_Wbar}.

The deviation originates from persistent accumulation of active particles near the piston boundaries, which modifies the empty region probability \(p_1\). This boundary layer produces two competing optimal regimes. At low area fraction, large detection regions maximize the output by storing more work per successful compression despite their lower probability. At higher area fractions, boundary vacancies are strongly suppressed, and the optimum shifts to smaller detection regions within the accumulation layer, operating at larger \(p_1\) but with lower work per step.

The same competition persists in the finite-time two-stroke cycle. There, performance is set not only by the extracted work, but also by the waiting time for a successful measurement and also by the dry-friction-dominated expansion dynamics. Thus, the boundary-controlled switch in optimal operation is not a quasistatic artifact, but survives under realistic cycle constraints.

Coarse-grained active-profile and two-density models capture the leading trends using only the measured boundary structure, showing that boundary accumulation explains much of the modified feedback statistics. Remaining discrepancies likely reflect correlations neglected in these descriptions, such as clustering, aggregation, alignment, or other collective effects within the active boundary layer.

More broadly, because boundary accumulation is widespread in confined persistent active matter, this mechanism should extend beyond the present bristle-robot realization. Active information engines act as fluctuation selectors: rather than harvesting bulk active forces directly, they extract work by selecting rare boundary configurations. This suggests a design principle in which boundary geometry, wall interactions, and confinement tune \(p_1\), control feedback statistics, and optimize finite-time performance. Active information engines must therefore be designed not only by their equation of state but also by the spatial statistics of the fluctuations selected by feedback.


\begin{acknowledgments}
\textit{acknowledgments -} Y.R., L.H., and R.G. acknowledge support from the European Research Council (ERC) under the European Union’s Horizon 2020 research and innovation program (Grant Agreement No. 101002392). N.B.A. and S.R. are grateful for support from the Israel Science Foundation (Grant No. 1929/21).
\end{acknowledgments}


\section*{End Matter}
\subsection*{Experimental details}
The bbots are approximately $4.0$\thinspace{cm} long and $1.2$\thinspace{cm} wide
($A_b=4.8\,\mathrm{cm^2}$). Each is powered by an LR44 battery driving
a rotational motor whose vibrations propel the bbot through twelve
flexible bristles. Experiments use $N_b=3$--$12$ bbots, with most
measurements performed at $N_b=6$ and the area fraction varied primarily
by changing $L_x$.

Each bbot carries a colored marker whose position is extracted from the recorded images using color-based tracking in MATLAB. Before measurement, the system is allowed to reach a stationary spatial distribution, typically after $\sim5$\thinspace{min}, as determined from the time evolution of the $x$-position histogram. One-dimensional density profiles are constructed from particles within a central strip of the arena, using the normalized coordinate $u=x/L_x$ and 150 spatial bins.

The correlation time used to set the measurement interval is obtained from the autocorrelation of a binary signal that is unity when the detection region is empty and zero otherwise. An exponential fit is performed for each $\Phi$ and $\Delta x$; the largest correlation time is approximately $1$\thinspace{s}, and we use $t_m=5\tau_{\rm corr}\simeq5$\thinspace{s}.

For the free-expansion measurements, the actuator-driven wall is replaced by a rigid 3D-printed PLA partition that slides directly on the arena floor. Several markers along the partition are tracked, and its position is defined as their mean $x$-coordinate.

\section*{Supplemental Materials}
	These Supplemental Materials provide additional details of the empty-region probability, details of the two-density model and the active profile model, deconvolution procedure, and finite-time expansion dynamics discussed in the main text.

\section{Direct estimate of the empty-region probability}

As a consistency check on the independent-particle approximation used to calculate $p_1$ from the single-bbot density profile, we also estimate the empty-region probability directly from the measured distribution of the leftmost bbot. In this approach,
\begin{equation}
	p_1 = P\left(x_{\mathrm{min}}>\Delta x\right),
\end{equation}
where $x_{\mathrm{min}}$ is the position of the leftmost bbot in a given frame.

\begin{figure}[h]
	\centering
	\includegraphics[width=1\columnwidth]{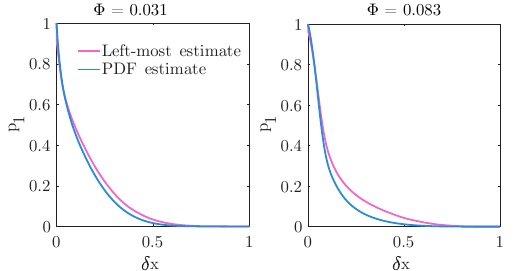}
	\caption{Comparison of the empty-region probability $p_1(\delta x)$ obtained from the normalized single-bbot density profile (PDF estimate) and directly from the measured distribution of the leftmost bbot, for two representative area fractions, $\Phi=0.031$ and $\Phi=0.083$.}
	\label{fig:leftmost_validation}
\end{figure}

Representative comparisons between the direct estimate and the density-profile-based estimate are shown in Fig.~\ref{fig:leftmost_validation}. The two methods show similar behavior over the measured range.

This direct estimate has the advantage that it does not rely on the assumption of independent bbot positions. However, it is statistically less efficient because each frame is reduced to a single extreme position rather than using the full set of bbot coordinates to estimate the underlying spatial distribution. As a result, the estimate is particularly sensitive to the limited number of steady-state configurations available in some datasets, especially when empty-region events are rare.

In addition, the leftmost-bbot statistic is intrinsically more sensitive to individual tracking errors or missed detections, since a single incorrectly identified extreme position can directly alter whether a frame is classified as empty or occupied. We therefore use the density-profile-based estimate for the quantitative analysis in the main text and retain the leftmost-bbot calculation as an independent consistency check.

\section{The two-density model}
\label{App:TwoDensity}

\begin{figure}[h!]
	\centering
	\includegraphics[width=0.9\linewidth]{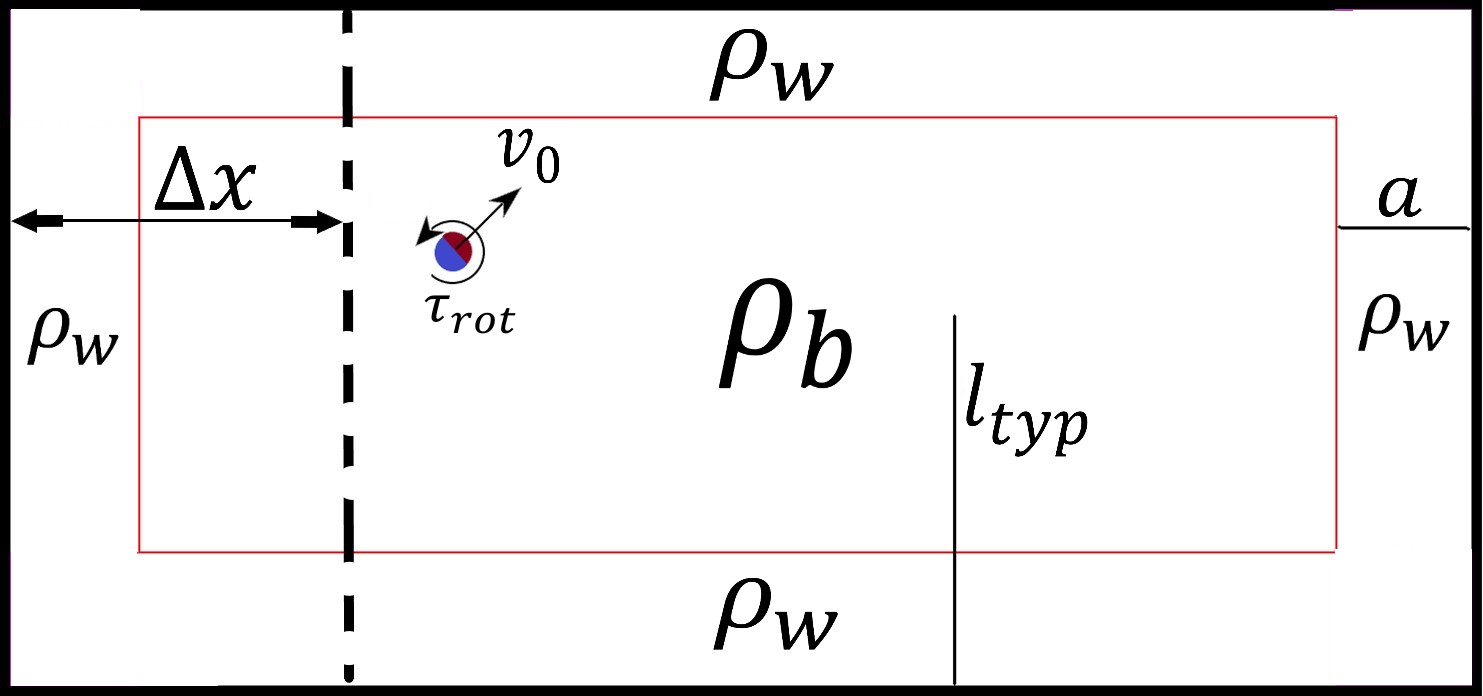}
	\caption{The two-density model. A high, but constant density of $\rho_w$ is assumed in a strip of width $a$ near the walls. A lower, but also constant density of $\rho_b$ is assumed in the bulk. }
	\label{fig:2_densities_model}
\end{figure}

As discussed in the main text and shown in Fig.~3(a) there, the particle density is high near the boundary, and low in the bulk. This suggests a simple two-density description of the particle distribution, expressing their tendency to be near the walls. In this description, one makes the simplified assumption of a high particle density $\rho_w$ in a strip of width $a$ near the boundary, and a much smaller particle density $\rho_b$ in the bulk of the system. To keep the description simple, both densities are assumed to be constant. The configuration is schematically depicted in Fig.~\ref{fig:2_densities_model}.
This simple description can then be used to calculate the thermodynamic and information properties of the engine, which are then compared to the experimental results. 

First, the two densities are related to a small set of parameters by demanding two consistency constraints. The particle probability density must be normalized. For a single particle this takes the form $n_w+n_b=\rho_w \left[L_x L_y - (L_x-2a)(L_y-2a) \right]+\rho_b (L_x-2a)(L_y-2a)=1$. ($n_{w,b}$ denote the number of particles near the wall, or in the bulk, respectively.)
In addition, the probability profile must be at steady state, meaning that the flux of particles leaving the boundary must be balanced by a flux of particles arriving from the bulk. The particle flux from the boundary to the bulk is estimated as $n_w/\tau_{rot}$. The reverse flux is $n_b v/l_{typ}$, where $v$ is the mean velocity and $l_{typ} \approx L_x/2,~L_y/2$ is the mean distance of a bulk particle from the wall.
From the consistency conditions, the two densities can be written as

\begin{equation} \label{eq:2D_densities_exact}
	\begin{split}
		\rho_w=&\frac{l_{per}}{2a(l_{typ}+l_{per})(L_y+L_x-2a)},\\
		\rho_b=&\frac{l_{typ}}{(l_{typ}+l_{per})(L_x-2a)(L_y-2a)}.
	\end{split}
\end{equation}

Here $l_{per}=v\tau_{rot}$ is the persistence length of a free bbot. The quality of this assumption can be estimated by comparing the dual density model to the measured projection of particle density  on the x-axis (Fig.~3(a) of main text).

It is straightforward to use eq. (\ref{eq:2D_densities_exact}) and calculate the probability $p_1$ that a strip of width $\Delta x$ near the boundary is empty. For a single particle, one finds
\begin{widetext}
	\begin{equation} \label{eq:prob_2fluid_models}
		\begin{gathered}
			p_1(\Delta x)
			= \begin{cases}
				L_y \left( L_x -\Delta x \right) \rho_w,  & L_x-a \leq \Delta x \leq L_x\\
				a \Big( L_y+2L_x-2a-2\Delta x \Big) \rho_w 
				+  \Big( L_y-2a\Big) \Big( L_x-a-\Delta x\Big) \rho_b, & a \leq \Delta x \leq L_x-a\\
				1-L_y \Delta x \rho_w, & 0 \leq \Delta x \leq a
			\end{cases}
		\end{gathered}
	\end{equation}
\end{widetext}
This can be generalized for more particles, as long as their density is low enough so that their interactions with each other can be neglected. In that case, one uses
$p_1(N)=p_1^N$, where $N$ is the number of active particles.

This simple description can also be used to estimate the work $W_{out}$ that can be extracted from a quasi-static expansion from $L_x-\Delta x$ back to $L_x$. The force on the wall is given by $F_1 \times L_y a \times \rho_w$, where $F_1\simeq 8.79$\thinspace{mN} is the mean force of one particle. A straightforward calculation results in
\begin{equation} \label{eq:work_2fluid_models}
	W_{out}=\frac{F_1 L_y l_{per}}{2(l_{typ}+l_{per})}\ln\left(\frac{L_y+L_x-2a}{L_y+L_x-2a-\Delta x}\right).
\end{equation}

The behavior of this simplified model is compared to experimental measurements in Fig.~3 of the main text. The comparison uses only a few parameters that need to be determined. $1.1 <a<1.9$\thinspace{cm} was determined by fitting the model density profile to the experimental PDF. $l_{per}=25\,\text{cm}$ is estimated from the trajectories of free bbot. Finally, $l_{typ}$ was taken to be $L_x/2$. Given the simplicity of the model and the fact that the few free parameters are determined from various calibration measurements, the results of the model match the experimental results remarkably well. Some deviations can be seen for $p_1 \rightarrow 1$, or $\Delta x \rightarrow 0$. These are expected since the finite size of the particle means that its center of mass coordinate cannot be arbitrarily close to the wall, forcing deviations from the assumption that $\rho_w$ is constant. Nevertheless, this simple description allows the interpretation of the two-region structure seen in Fig.~3(d) as a result of $\Delta x$ values that are in the boundary layer or in the bulk.

\section{Derivation of the active-profile model}
\label{sec:active_profile_model}

The active-profile model describes the measured accumulation of bbots near the boundaries using a bulk density and exponentially decaying high-density layers. We describe the position of a single bbot by the normalized density profile
\begin{equation}
	\rho(x)=\rho_b+\rho_{\mathrm{ex}}
	\left[
	e^{-(x-\ell_{\mathrm{eff}})/\lambda}
	+
	e^{-(L_x-\ell_{\mathrm{eff}}-x)/\lambda}
	\right],
	\label{eq:active_profile}
\end{equation}
for $\ell_{\mathrm{eff}}\leq x\leq L_x-\ell_{\mathrm{eff}}$, and $\rho(x)=0$ otherwise. Here $\rho_b$ is the bulk density, $\rho_{\mathrm{ex}}=\rho_w-\rho_b$ is the excess near the walls, $\lambda$ is the decay length of the high-density layer, and $\ell_{\mathrm{eff}}$ accounts for the finite size of the bbots.

Defining the accessible system length as
\begin{equation}
	L_{\mathrm{acc}}=L_x-2\ell_{\mathrm{eff}},
\end{equation}
the normalization condition becomes
\begin{equation}
	1=
	\rho_bL_{\mathrm{acc}}
	+
	2\rho_{\mathrm{ex}}\lambda
	\left[
	1-e^{-L_{\mathrm{acc}}/\lambda}
	\right],
	\label{eq:active_profile_normalization}
\end{equation}
and therefore solving for the excess density gives
\begin{equation}
	\rho_{\mathrm{ex}}=
	\frac{1-\rho_bL_{\mathrm{acc}}}
	{2\lambda\left[1-e^{-L_{\mathrm{acc}}/\lambda}\right]}.
	\label{eq:rho_ex_active_profile}
\end{equation}

For each area fraction, $\rho_b$ is obtained from the central region of the measured density profile. The decay length $\lambda$ is determined by scanning over possible values and minimizing the Kullback--Leibler divergence between the model and the measured profile. In the calculations shown in the main text, $\ell_{\mathrm{eff}}\simeq1.2$\thinspace{cm}, as obtained from the deconvolution procedure described in Sec.~\ref{sec:pdf_deconvolution}.

We next calculate the probability $p_1(\Delta x)$ that the detection region of width $\Delta x$ adjacent to the piston is empty. Since $\rho(x)$ is normalized to unity, the probability that a single bbot lies within the detection region is
\begin{equation}
	q(\Delta x)=
	\int_{\ell_{\mathrm{eff}}}^{\ell_{\mathrm{eff}}+\Delta x}
	\rho(x)\,dx .
	\label{eq:q_active_profile}
\end{equation}
For $\Delta x\leq L_{\mathrm{acc}}$, this gives
\begin{widetext}
	\begin{equation}
		q(\Delta x)=
		\rho_b\Delta x
		+
		\rho_{\mathrm{ex}}\lambda
		\left[1-e^{-\Delta x/\lambda}\right]
		+
		\rho_{\mathrm{ex}}\lambda
		\left[
		e^{-(L_{\mathrm{acc}}-\Delta x)/\lambda}
		-e^{-L_{\mathrm{acc}}/\lambda}
		\right].
		\label{eq:q_active_profile_explicit}
	\end{equation}
\end{widetext}

Neglecting correlations between the positions of different bbots, the probability that none of the $N_b$ bbots lies within the detection region is
\begin{equation}
	p_1(\Delta x)=
	\left[1-q(\Delta x)\right]^{N_b}.
	\label{eq:p1_active_profile}
\end{equation}
This expression gives the active-profile prediction shown by the blue curves in the main text.

In the homogeneous limit, $\rho_{\mathrm{ex}}\rightarrow0$ and $\ell_{\mathrm{eff}}\rightarrow0$, the normalized density becomes $\rho(x)=1/L_x$, so that
\begin{equation}
	q(\Delta x)=\frac{\Delta x}{L_x}=\delta x,
\end{equation}
and the ideal-gas result is recovered.

\section{Constrained deconvolution of the measured density profile}
\label{sec:pdf_deconvolution}

The measured one-dimensional density profile is obtained from the tracked sticker positions along the long axis of the arena. Because the sticker is not always located exactly at the bbot center, and because the bbots are elongated, the measured coordinate can differ systematically from the effective center coordinate. This produces an apparent nonzero probability density close to the walls, including regions that are inaccessible to the bbot center.
\begin{figure}[h]
	\centering
	\includegraphics[width=1\linewidth]{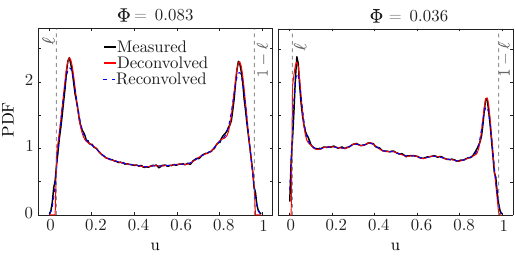}
	\caption{Validation of the density-profile deconvolution for two representative area fractions, $\Phi=0.083$ and $\Phi=0.036$. The measured positional density profile (black), deconvolved profile $\rho_{\rm true}$ (red), and reconvolved profile $\rho_{\rm reconv}$ (blue dashed) are shown as functions of the normalized
		position $u=x/L_x$. The vertical dashed lines indicate the boundaries of the physically accessible region, $u=\ell$ and $u=1-\ell$, with
		$\ell=\ell_{\rm eff}/L_x$. The close agreement between the measured and reconvolved profiles demonstrates that the constrained deconvolution
		reproduces the experimental distribution while enforcing the geometric exclusion of bbot centers near the walls.}
	\label{fig:deconv_validation}
\end{figure}
We model the measured coordinate as
\begin{equation}
	x_{\rm meas}=x_{\rm true}+\eta,
\end{equation}
where the effective localization error $\eta$ is approximated as Gaussian with width $\sigma_x$. Using the normalized coordinate $u=x/L_x$
and $\sigma=\sigma_x/L_x$, the measured and true density profiles are related by
\begin{equation}
	\rho_{\rm meas}(u)
	=
	\int_{\ell}^{1-\ell}
	G_{\sigma}(u-u')\rho_{\rm true}(u')\,du',
	\label{eq:measured_convolution_allowed}
\end{equation}
where
\begin{equation}
	G_{\sigma}(u-u')
	=
	\frac{1}{\sqrt{2\pi}\sigma}
	\exp\left[-\frac{(u-u')^2}{2\sigma^2}\right],
\end{equation}
and
\begin{equation}
	\ell=\frac{\ell_{\rm eff}}{L_x}.
\end{equation}
The effective excluded length $\ell_{\rm eff}$ accounts for the finite size of the bbots, so that
\begin{equation}
	\rho_{\rm true}(u)=0
	\qquad
	\text{for}
	\qquad
	u<\ell
	\quad \text{or} \quad
	u>1-\ell .
\end{equation}

The convolution is discretized on the same grid used for the measured profile,
\begin{equation}
	\boldsymbol{\rho}_{\rm meas}
	\simeq
	K_{\mathcal A}\mathbf r,
\end{equation}
where $K_{\mathcal A}$ is the Gaussian convolution matrix restricted to the physically accessible interval and $\mathbf r$ contains the unknown values of the deconvolved profile within that interval.

Because deconvolution amplifies noise in the measured density profile, we include a smoothness regularization term to suppress unphysical high-frequency oscillations in the reconstructed profile:
\begin{equation}
	\mathbf r^{\ast}
	=
	\underset{\mathbf r}{\operatorname{argmin}}
	\left[
	\left\|
	K_{\mathcal A}\mathbf r
	-
	\boldsymbol{\rho}_{\rm meas}
	\right\|_2^2
	+
	\lambda_{\rm reg}
	\left\|
	D_2\mathbf r
	\right\|_2^2
	\right],
	\label{eq:regularized_deconvolution}
\end{equation}
subject to
\begin{equation}
	r_j\geq0,
	\qquad
	\sum_j r_j\Delta u=1.
\end{equation}
Here $D_2$ is the discrete second-derivative operator and $\lambda_{\rm reg}$ controls the strength of the smoothness constraint. This procedure suppresses oscillatory artifacts while preserving the overall structure of the density profile without assuming a specific functional form.

The effective excluded length $\ell_{\rm eff}$, localization width $\sigma$, and regularization strength $\lambda_{\rm reg}$ are selected by scanning over physically reasonable values and comparing the reconvolved profile with the measured one. For each candidate solution, the reconstructed profile is convolved again
with the same Gaussian kernel,
\begin{equation}
	\boldsymbol{\rho}_{\rm reconv}
	=
	K\boldsymbol{\rho}_{\rm true}.
\end{equation}

Only solutions that remain non-negative, normalized, confined to the accessible interval, and reproduce the measured profile are kept. Representative examples are shown in Fig.~\ref{fig:deconv_validation}.

The resulting deconvolved profile $\rho_{\rm true}(u)$, shown in Fig.~\ref{fig:deconv_validation}, is used throughout the analysis to determine the single-bbot probability of occupying the detection region and, from it, the empty-region probability $p_1(\Delta x)$.

\section{Simulations of expansion with dry friction}

Simulations of a simple model were used to test whether the expansion dynamics shown in Fig.~4(b) of the main text can be attributed to a stick-slip mechanism. The simulations can also reveal whether the qualitative behavior observed there is robust, or if it is seen only for some parameters.

Our model of the expansion process assumes stick-slip motion with Coulomb dry friction. The piston location and mass are denoted by $X$ and $M$ respectively. When the piston is moving, $\left| \dot{X} \right|>0$, its dynamics evolve according to
\begin{equation}
	M \ddot{X} = F(n(t))-F_k \operatorname{sgn}(\dot{X}) - F_e.
	\label{eq:newtonfric}
\end{equation}
Here $F(n(t))$ is the force applied by the bbots, $F_k$ is the dynamical friction, $\operatorname{sgn}$ is the sign function, and $F_e$ denotes an externally applied load.
Since all the bbots are confined inside the piston, they always push it outwards, such that $F(n)\ge 0$. Similarly, the load always resists the piston motion, so $F_e\ge 0$.

Once the piston stops, it stays at rest until the net force on it overcomes a static friction force, $\left| F(n)-F_e \right|>F_s$. When this threshold is crossed, the piston continues to evolve according to Eq.~(\ref{eq:newtonfric}). We assume that $F_s>F_k$.

We use a very simple model for the bbot dynamics. We denote the total number of bbots in the system by $N$. At any time, $n(t)$ of them are pushing the piston, while $N-n(t)$ are along the opposite wall. We assume that the force on the piston is
\begin{equation}
	F(n) = f_b n(t),
\end{equation}
namely, we use the mean force of each bbot, rather than a more detailed description.

The number $n(t)$ follows a Markov jump process in which the possible processes are either i) $n \rightarrow n+1$, due to an arrival of a bbot to the piston, or ii) $n \rightarrow n-1$, due to a bbot that leaves the piston region.
The relevant transition rates have the form
\begin{equation}
	r_n^{(+)} = \frac{(N-n)r^{(+)}}{\left| X(t)-X_0\right|},
\end{equation}
and
\begin{equation}
	r_n^{(-)} = n r^{(-)}.
	\label{eq:leave_rate}
\end{equation}
Both rates are based on the assumption that each bbot moves independently. However, the bbot arrival rate is inversely proportional to the arena area, and thus also to the piston's distance from the opposing wall. In contrast, the typical time for bbot escape from the wall is roughly its rotation time, which is independent of $X(t)$. Note that the rates ensure that $ 0 \le n(t) \le N$.

This simple model exhibits several physically appealing characteristics, which may be absent in models that use Gaussian noise for the stochastic component of the force. The changes in the force $F(n)$ are limited to the force of a single bbot. Models with white or short-memory noise exhibit larger jumps in the force. Moreover, this model ensures that $F(n(t)) \ge 0$,
which is consistent with the presence of bbots on one side of the piston. Modeling the stochastic force with Gaussian noise will inevitably result in wrong signs, even if these instances are rare events.

\begin{figure}[h!]
	\centering
	\includegraphics[width=1\linewidth]{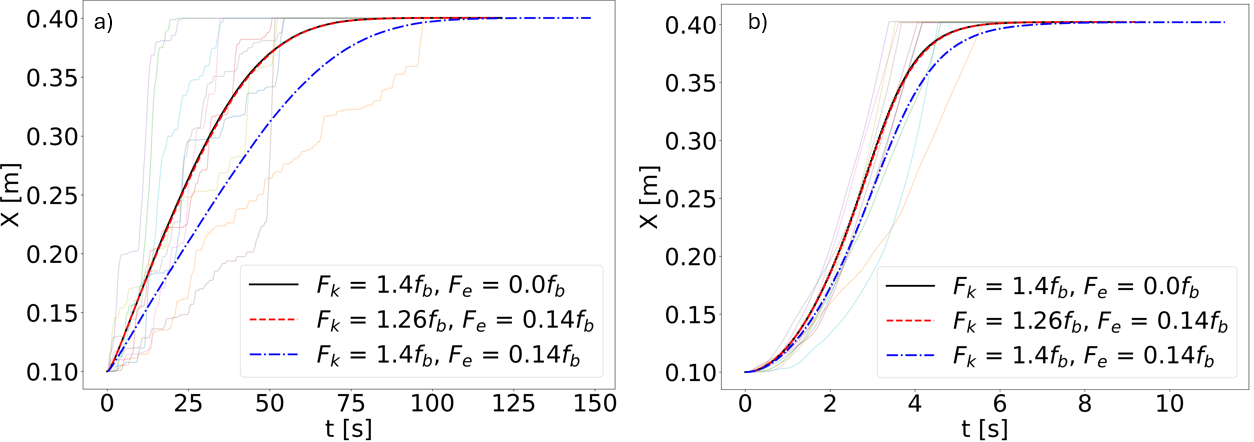}
	\caption{Simulation results of expansion with dry friction. In both panels - $N=6$, $M=3.05$\thinspace{g}, $f_b=1.689$\thinspace{mN} and $r^{(-)}=10$\thinspace{Hz}. Panel (a) has $F_s=2.5f_b$, and $r^{(+)}=0.55$\thinspace{Hz} whereas panel (b) has $F_s=1.5f_b$ and $r^{(+)}=1.55$\thinspace{Hz}.}
	\label{fig:simulated_stick_slip}
\end{figure}

The coupled dynamics of the piston and bbots was simulated using Eqs.~(\ref{eq:newtonfric}) - (\ref{eq:leave_rate}), and the results are depicted in Fig.~\ref{fig:simulated_stick_slip}. 
The thin gray lines in the panels are traces of single realizations of the simulation. The thick lines are ensemble averages over 10000 repetitions. The results in panel (a) show strong similarity to measurements depicted in Fig.~4(b) of the main text.

The results in panel (a) exhibit stick-slip dynamics, due to the need for at least $3$ bbots to overcome the static friction. The results in panel (b) were obtained for lower static friction and a larger value of $r^{(+)}$. As a result, a qualitatively different dynamics, with significantly fewer sticking events, is found.

The panels also include ensemble averages of simulations with a resisting force. Thick black lines correspond to runs with only friction force. The blue dashed-dotted line is the mean of runs with an additional external load. The dashed-red line is from runs with the additional load, but with reduced dynamic friction so that the sum of the forces is equal to the friction force used in the other runs. 

The results from both panels show that adding a small load leads to a qualitatively similar, but quantitatively slower expansion process. Moreover, the expansion with the load and smaller friction is identical to that obtained with only friction. This is not surprising, since the force that the piston feels during its expansion is tuned to be the same in both cases. The identical curves suggest that the measured expansion curve depicted in Fig.~4(b) of the main text, which was obtained without an external load, would also describe a realistic cycle, with a load and somewhat lower friction.
\end{document}